\documentclass[epj]{svjour}
\usepackage{graphics,amsmath}

\title{Comparing Mean Field and Euclidean Matching Problems}

\author{J. Houdayer\inst{1}\and J. H. Boutet de Monvel\inst{2}\fnmsep\inst{1}\and
O. C. Martin\inst{1}}

\institute{
Division de Physique Th\'eorique, Institut de Physique Nucl\'eaire,
Universit\'e Paris-Sud, F--91406 Orsay Cedex,
France.\\
\email{houdayer@ipno.in2p3.fr, martino@ipno.in2p3.fr}\and
Forschungszentrum BiBos, Fakult\H{a}t f\H{u}r Physik, Universit\H{a}t
Bielefeld, D--33615 Bielefeld, Germany.\\
\email{Jacques.Boutet.de.Monvel@ood.ki.se}}

\offprints{J. Houdayer}

\mail{J. Houdayer}

\date{\today}

\abstract{Combinatorial optimization is a fertile testing ground for 
statistical physics methods developed in the 
context of disordered systems, allowing
one to confront theoretical mean field predictions with actual
properties of finite dimensional systems. Our focus here is on minimum
matching problems, because they are computationally tractable while both
frustrated and disordered. We first study a mean field model taking the link
lengths between points to be independent random variables. For this model we
find perfect agreement with the results of a replica calculation, and
give a conjecture. Then we
study the case where the points to be matched are placed at random in a
$d$-dimensional Euclidean space.
Using the mean field model as an approximation to the Euclidean case,
we show numerically that the
mean field predictions are very accurate even at low dimension, and that the
error due to the approximation is $O(1/d^2)$. Furthermore, 
it is possible to improve upon this approximation 
by including the effects of Euclidean correlations among 
$k$ link lengths.  Using $k=3$ (3-link correlations such as the triangle
inequality), the resulting errors in the energy density are already
less than $0.5\%$ at $d\ge2$. However, we argue that 
the dimensional dependence of the Euclidean model's energy density
is non-perturbative, {\it i.e.}, it is beyond all orders
in $k$ of the expansion in $k$-link correlations.
}

\begin{document}
\sloppy
\maketitle

\section{Introduction}
\label{SectIntro}

\subsection{Background}
\label{SectBckgrnd}
The study of disordered and frustrated systems, and in particular
spin-glasses, has long been a major issue in condensed matter physics (for
reviews see \cite{MezardParisi87b,BinderYoung86,FischerHertz91}). Most
efforts have been based on replicas, in part because that method has led to
the exact solution \cite{Parisi80,Parisi80b} of the Sherrington-Kirkpatrick
(SK) model~\cite{SherringtonKirkpatrick75}. However, since the SK model is of
infinite range, it is not 
clear \cite{BrayMoore84,FisherSompolinsky85,FisherHuse88}
how relevant its solution is for understanding finite dimensional
spin-glasses such as the Edwards-Anderson (EA)
model~\cite{EdwardsAnderson75}. The application of the replica formalism to
finite dimensional systems, on the other hand,
is hampered by two major difficulties
\cite{GeorgesMezard90,DominicisKondor91}. First, in the saddle point
equations, finite connectivities lead to an infinite number of
order parameters: one has to deal with order parameters
$q_{\alpha\beta\gamma\ldots}$ having an arbitrarily large number of
indices~\cite{DominicisMottishaw87}. In contrast, the SK model requires
only the order parameter $q_{\alpha\beta}$, with two
indices. The second difficulty comes from the Euclidean nature of space:
The metric structure 
introduces constraints on the possible values of the quenched disorder
variables between points $i$ and $j$. In the infinite range model, these
variables are independent, but in the short range models, the values allowed
depend on the distance between points $i$ and $j$, thereby introducing strong
correlations. These correlations make the replica analysis much more
difficult.

In this article we study matching problems; these are
disordered and frustrated short range models arising
in combinatorial optimization. They are simpler than spin-glasses and
the difficulties just stressed have to a large extent been overcome.
In particular, for the minimum matching problem based on {\it independent
random} link
lengths, M\'ezard and Parisi (M \& P) have worked out
the order parameters $q_{\alpha\beta\gamma\ldots}$
exactly~\cite{MezardParisi85}
and have introduced a link correlation expansion~\cite{MezardParisi88}
to take into account correlations among link
variables. Using extensive numerical simulations of the ground states,
we study the following: (i) the probability distribution of ground
state energies in the disorder ensembles; (ii) the distribution
of ``local'' energies in the ground state configurations; 
(iii) the validity of their replica approach in the
case without correlations; (iv) the accuracy 
of this random-link length model as a mean-field
approximation to the Euclidean model; (v) the accuracy of the
link correlation expansion.

\subsection{The models}
\label{SectModel}
Consider $N$ points ($N$ even), and a set of links with lengths
$l_{ij}=l_{ji}$ connecting points $i$ and $j$, ($i,j=1,\ldots,N$). We
call an {\em instance} a specification of these link lengths. A {\em
matching} of the points is a set of bonds in which each of the $N$ points is
the extremity of one and only one bond. In effect this is a dimerization:
points are linked in pairs. The length or cost
of a matching is then the sum of the
lengths of its bonds. (Since the points will often
belong to a Euclidean space, we will use the term length rather
than cost hereafter.) The minimum matching problem
(MMP)~\cite{PapadimitriouSteiglitz82} can be defined as the problem of
finding a matching of minimum length, given the $l_{ij}$'s. A variant of
this problem is the minimum bipartite matching problem
(MBMP)~\cite{PapadimitriouSteiglitz82}: we now have {\it two} sets of $N$ points
instead of one ($N$ is no longer necessarily even) and a bipartite matching
consists of a dimerization where each pair contains one point from each set.
The length and the minimization of the bipartite
matching are defined as for the MMP.  From
an algorithmic point of view, both problems belong to the class $P$,
meaning that the exact solution of any
instance of size $N$ can be found in a
time at most polynomial in $N$.  In fact, the MMP can be solved in a time of
$O(N^3)$ and the MBMP in a time of $O(N^2\ln N)$. This property is
important for a numerical study since it allows
extensive exact computations. We have
implemented the $O(N E \ln(E))$ matching algorithm
as exposed in \cite{BallDerigs83} where
$E$ is the number of edges in the graph. The resulting program
solves a typical $N=200$ random point instance in less than a
second on a Dec-Alpha machine. The total computation time spent 
generating the data summarized in the tables of this paper 
amounts to about $100$ days of machine time.

These combinatorial optimization problems may be mapped onto physical
systems, where each matching is one state of the system with an energy equal to
its length. Then the minimum matching problem is equivalent to finding the
ground state of that system. The physical systems built in this way are
frustrated since in general not all points may be matched with their nearest
neighbor.  The thermodynamics of these systems can be studied by
introducing Boltzmann weights for the states as was proposed by Kirkpatrick
{\em et al.}~\cite{KirkpatrickGelatt83}. Since this article focuses on ground
states, however, we restrict our discussion to the zero temperature properties
of these systems.

In any disordered system, be it a spin-glass or a matching problem, one is
not particularly interested in the properties of a given instance of the
problem. More relevant physically are typical properties, or averages over
an ensemble of instances. The $l_{ij}$'s then become quenched random
variables and one speaks of the stochastic M(B)MP, problems that are both
disordered and frustrated. Moreover, we are interested in the infinite
``volume'' limit, meaning in this case the limit $N\rightarrow\infty$. These
systems can then be studied using the replica or the cavity method (for a
review see~\cite{MezardParisi87b}) as developed by M \& P
\cite{MezardParisi85,MezardParisi86}.

Let us now describe the different ensembles of $l_{ij}$'s we consider. The first
is the Euclidean ensemble (for the MMP as well as for the MBMP) where we
have $N$ (or $2N$) random and independent points chosen uniformly in a
$d$-dimensional volume (e.g., a unit hypercube), and the lengths $l_{ij}$
are given by the Euclidean distances between the points. This is a
short-range model in that the points lie in a Euclidean space and only the
first few nearest neighbors are relevant for the minimum matching. 
(In the formulation based on spins \cite{MezardParisi85},
the coupling between spins decreases exponentially with distance.) Using a
spin-glass analogy, this Euclidean model is like the EA model (both models
are short range, leading to important Euclidean correlations in the disorder
variables). Not surprisingly, the Euclidean MMP --- like the EA model --- is not
solvable analytically. This suggests that one should consider a different
ensemble for the disorder variables in order to render the problem more
tractable~\cite{VannimenusMezard84}. Indeed, this may be accomplished
by taking the lengths $l_{ij}$ to be independent, identically
distributed random variables. The corresponding model is called the
``random-link'' model. (Note that the points do not lie in a metric space
and the $l_{ij}$'s no longer have correlations; in particular the
triangle inequality does not hold). Pursuing our spin-glass analogy, the
random-link MMP is like the model of Viana and Bray~\cite{VianaBray85} for
spin-glasses.  Both models are ``infinite dimensional'' in the 
sense that
there is no underlying geometry and thus there are no Euclidean
correlations. Furthermore, in both cases, the 
effective connectivity at each site
stays fixed as the size of the system grows. For the Viana and Bray model,
this is enforced by having the number of non-zero couplings to a spin be
size independent; for the random-link MMP, this occurs because
only the first few nearest neighbors of a site effectively
contribute to the minimum
matching. These models thus interpolate between the infinite range /
infinite connectivity case and the finite range / finite connectivity case.
Thus, since it is expected that the Viana and Bray model provides a better
approximation than the SK model to the EA model, the fact that its analogue
here (the random-link MMP) is exactly solvable is of major interest.

A connection between the Euclidean and the random-link models was first
given by M \& P \cite{MezardParisi86}: they pointed out that the one- and
two-link distributions could be made identical in both problems,
leading to a particular family of random-link models
parametrized by $d$, where $d$ is the spatial dimension for the
corresponding Euclidean model. With such a
choice, both models have the same Cayley tree approximation. One may then
consider the random-link MMP to be a mean-field model for the Euclidean MMP
in which correlations between link lengths (the quenched disorder variables)
have been neglected. What we call from now on the ``random-link
approximation''~\cite{CerfBoutet97} consists of using the thermodynamic
functions of the random-link models as estimators for those of the Euclidean
models.

\subsection{Outline}
\label{SectOutln}
This paper expands upon previous work~\cite{BoutetMartin97},
and provides an in depth study of ground state properties
(disorder induced distributions of the energy, length of dimers, 
dimensional dependence, etc...) in the MMP and the MBMP.
The outline is as
follows. In section~\ref{SectRLMdl}, we examine the random-link models. 
First, we discuss self-averaging and large $N$ scaling properties of the ground
state energy. Second, we recall results derived assuming
replica symmetry and check them numerically.
Third, we propose a conjecture for the frequency with which a point
connects to its $k$th nearest neighbor in the ground state of the M(B)MP. In
section~\ref{EUModels}, we study the Euclidean models. First, we discuss
self-averaging of the ground state energy. Second, we quantify numerically
the precision of the random-link approximation. Third, we consider the
Euclidean corrections to this approximation in the case of the MMP. 
Finally,
section~\ref{Discussion} discusses our results and provides an outlook
on possible generalizations to other systems.

In Appendix~\ref{Bound}, we present an $N$-independent upper bound for the
length of the random-link MMP when $d=1$. In Appendix~\ref{StrongSA}, we prove
a self-averaging property for the Euclidean MBMP in dimensions
greater or equal to 3.

\section{Random-link models}
\label{SectRLMdl}
In order to use the random-link models as an approximation for the Euclidean
ones, we must, as previously mentioned,
set the random-link distribution to match
the one-link Euclidean distribution. 
In the Euclidean model, one possible approach would be to take the
large $N$ limit at fixed density of points, in which case the
volume would scale linearly with $N$. However, for historical 
reasons, the standard practice is to take $N \to \infty$
in a fixed volume. These two pictures are equivalent, and are 
mapped onto one another by rescaling all the lengths by a factor
$N^{1/d}$. As a consequence, in the units we shall use in the rest 
of this paper,
the mean length between
neighboring points scales as $N^{-1/d}$. 

Consider two points $i$ and $j$
randomly chosen in the $d$-dimensional unit hypercube. In the
absence of edge effects, the distribution of
their distance $l_{ij}$ is given by
$\rho_d(l_{ij}=r)=dB_dr^{d-1}$, where $B_d=\pi^{d/2}/(d/2)!$ is the volume
of the $d$-dimensional unit ball. We thus take from now on $\rho_d(l)$ as
the individual distribution for the link lengths in the random-link model.
The random-link M(B)MP models are then described by a single parameter $d$.
Two comments are in order. First, $\rho_d$ is not normalized, and
so must be cut off; this can be done arbitrarily as the
large $N$ scaling of the minimum matching
depends only on the behavior of 
$\rho_d(l)$ at small $l$. Second, neglecting edge effects, 
any two lengths are uncorrelated in Euclidean space, so the 
prescription just given also matches the two-link Euclidean distribution.

\subsection{The large $N$ limit}
\label{SectRLLargeN}
We denote by $L_{MM}^{RL}$ the length (or energy) of the minimum matching
in the MMP; $L_{MM}^{RL}$ is a random variable depending on the instance
({\it i.e.}, on the $l_{ij}$'s).  It is a sum of $N/2$ terms, each of which is
typically the length between near neighbors.
We have seen that these lengths scale
as $N^{-1/d}$, so $L_{MM}^{RL}$ is expected to scale as $N^{1-1/d}$. 
For instance, consider the length $d_i$ between point $i$ and
its nearest neighbor. It is easy to show that in the large $N$ limit, 
$\langle d_i \rangle \sim D_1(d) / N^{1/d}$ where
\begin{equation}
\label{eq_D_1}
D_1(d)=(1/d)!B_d^{-1/d}
\end{equation}
is the rescaled average nearest neighbor length at large $N$.
(In the units where the density of points is equal to one,
$D_1(d)$ is exactly the mean nearest neighbor length.)
In the case of the minimum matching length,
it can be proven~\cite{Boutet96} that $L_{MM}^{RL}/N^{1-1/d}$ becomes peaked
around its mean value as $N\rightarrow\infty$. A stronger property, called
self-averaging, would be that $L_{MM}^{RL}/N^{1-1/d}$ tends with probability
one to a (non-random, $N$-independent) constant $\beta_{MM}^{RL}(d)$, as
$N\rightarrow\infty$. Although this second property has not yet been proven,
it is strongly supported by previous numerical
studies~\cite{BrunettiKrauth91} as well as the simulations described in 
paragraph~\ref{SectRLNum}. Moreover, bounds can be 
found for $\beta_{MM}^{RL}(d)$ (see
Appendix~\ref{Bound}) that reinforce this hypothesis. In the following, we
assume the existence of the $\beta_{MM}^{RL}(d)$. The analogous discussion
applies to MBMP as well, so we will assume $\displaystyle\beta_{MBM}^{RL}(d)
=\lim_{N\rightarrow\infty}L_{MBM}^{RL}/N^{1-1/d}$.

It is of interest to understand the {\it distribution} of $L_{MM}^{RL}$ in the
large $N$ limit. In particular, one may wonder whether the self-averaging
property comes from some kind of central limit theorem which would lead to a
Gaussian (normal) limit distribution for $L_{MM}^{RL}$. The central 
limit theorem
(CLT) states that for a sum $S$ of $N$ independent identically
distributed random variables the
relative standard deviation $\sigma=\sqrt{Var(S)}/\langle S\rangle$ and the
skewness $s=\langle(S-\langle S\rangle)^3\rangle/\sigma^3$ decrease as
$1/\sqrt{N}$. Although in fact
the terms entering the sum $L_{MM}^{RL}$ are
correlated, if the correlations are not too strong, we can expect the same
CLT-type scaling to hold for $L_{MM}^{RL}$. We have computed $\sigma$ and $s$
numerically for $L_{MM}^{RL}$ and find a behavior
that is in excellent agreement with the
expected CLT scaling laws (see Table~\ref{TableGaussRL}).
(Since the finite size effects and the statistical noise
are significant, we have not extrapolated our data to the
$N \to \infty$ limit.)

\begin{table}
\caption{Numerical measurements of the relative standard deviation $\sigma$
and skewness $s$ of the distribution of $L_{MM}^{RL}$ and $L_{MBM}^{RL}$
at $d=1$ and $d=2$.}
\label{TableGaussRL}
\begin{center}
\begin{tabular}{|c|cc|cc|}
\hline\noalign{\smallskip}
& \multicolumn{2}{c|}{MMP $d=1$} & \multicolumn{2}{c|}{MMP $d=2$}\\
$N$ & $\sigma\sqrt{N/2}$ & $s\sqrt{N/2}$ & $\sigma\sqrt{N/2}$ & $s\sqrt{N/2}$\\
\noalign{\smallskip}\hline\noalign{\smallskip}
50 & 0.784 & 1.39 & 0.421 & 0.33\\
100 & 0.798 & 1.50 & 0.423 & 0.32\\
200 & 0.798 & 1.39 & 0.427 & 0.36\\
400 & 0.808 & 1.56 & 0.424 & 0.24\\
\noalign{\smallskip}\hline\hline\noalign{\smallskip}
& \multicolumn{2}{c|}{MBMP $d=1$} & \multicolumn{2}{c|}{MBMP $d=2$}\\
$N$ & $\sigma\sqrt{N}$ & $s\sqrt{N}$ & $\sigma\sqrt{N}$ & $s\sqrt{N}$\\
\noalign{\smallskip}\hline\noalign{\smallskip}
25 & 0.780 & 1.29 & 0.419 & 0.25\\
50 & 0.795 & 1.36 & 0.423 & 0.29\\
100 & 0.802 & 1.36 & 0.427 & 0.35\\
200 & 0.811 & 1.57 & 0.428 & 0.45\\
\noalign{\smallskip}\hline
\end{tabular}
\end{center}
\end{table}

Similarly, we have computed $\sigma$ and $s$ for the MBMP, and find nearly
identical results (see again Table~\ref{TableGaussRL}).  Clearly, the
large $N$ behavior follows the CLT scaling laws for both 
the MMP and the MBMP.
Furthermore, our data 
suggest that the amplitudes associated with these
scalings are equal in the two models. We conjecture that the two are
equal for {\it any} of the connected moments, and provide 
theoretical support for
this in section~\ref{SectAnltcal}. We will also derive a theoretical estimate
for the values of $\sigma$ and $s$ in section~\ref{SectRLNum}.

\subsection{Survey of analytical results}
\label{SectAnltcal}
The random-link MMP was first solved by M \& P \cite{MezardParisi85} using
the replica method, with a replica symmetric ansatz. They
verified~\cite{MezardParisi87} the stability of the replica symmetric
solution (at least for $d=1$), suggesting that most likely the ansatz is
exact. They also confirmed their results by the cavity 
method~\cite{MezardParisi86}. In our notation their results may be written
\begin{equation}
\beta_{MM}^{RL}(d)=\frac{dD_1(d)}{2(1/d)!}
\int_{-\infty}^{+\infty}G_d(x)e^{-G_d(x)}dx,
\end{equation}
where $G_d(x)$ satisfies the integral equation
\begin{equation}
\label{EquInt}
G_d(x)=d\int_{-x}^{+\infty}(x+y)^{d-1}e^{-G_d(y)}dy
\end{equation}
and $D_1(d)$ is as defined in equation~\ref{eq_D_1}.
Equation~\ref{EquInt} can be solved at $d=1$, leading to
$G_1(x)=\ln(1+e^x)$ and $\beta_{MM}^{RL}(1)=\pi^2/24$. Furthermore, M \& P
calculated the distribution $P_d(l)$ of the rescaled bond length 
$N^{1/d} l_{ij}$ in the
minimum matching in the limit $N\rightarrow\infty$, and found
\begin{equation}
P_d(l)=d\,l^{d-1}\int_{-\infty}^{+\infty}\frac{d G_d}{d x}(x)
e^{-G_d(x)-G_d(l-x)}dx.
\label{EquPl}
\end{equation}
In the case $d=1$, they found $P_1(l)=2(2l-e^{-2l}\sinh (2l))/\sinh^2 (2l)$.
Finally, since replica symmetry is not broken, one expects 
the mean length to have a $1/N$ expansion:
\begin{equation}
\frac{\langle L_{MM}^{RL}\rangle}{N^{1-1/d}}=
\beta_{MM}^{RL}(d)\left(1+\frac{A(d)}N+\frac{B(d)}{N^2}+\cdots\right).
\label{EquNDev}
\end{equation}
M \& P have calculated~\cite{MezardParisi87} the first subleading term at
$d=1$ and have found $A(1)\approx-0.13$.

Similar calculations were performed for the 
MBMP (Orland~\cite{Orland85}, 
M \& P \cite{MezardParisi85,MezardParisi86,MezardParisi87}). In our 
units, they find:
\begin{equation}
\beta_{MBM}^{RL}(d)=2\beta_{MM}^{RL}(d).
\label{EquMBMP}
\end{equation}
At $d=1$, they obtain $\beta_{MBM}^{RL}(1)=\pi^2/12$ and
$A(1) \approx -1.61$.

The MBMP and the MMP are very closely related. This is best seen within the
convention we have used where the MMP matches $N$ points, the MBMP matches
$2N$ points and both models have the same individual link length
distribution $\rho_d$. One then sees that the saddle point equations for
the partition functions (in the limit of large $N$) become identical in the
two models. Thus $P_d(l)$ is the same in the two models, and in fact any
reasonable observable will be the same in both models at large $N$. This
remarkable property has apparently gone unnoticed so far. 
(Given this property and our conventions, the factor $2$ in 
equation~\ref{EquMBMP}
simply follows from the fact that a bipartite matching
has twice as many bonds as a simple matching.) A consequence of
this correspondence is that the moments of $L^{RL}_{MM}$ and
$L^{RL}_{MBM}$ should be the same at large $N$. We are thus able to provide
theoretical support for the conjecture, given in the previous section,
and based on our numerical results.

It is also of interest to study the limit of large $d$. We have derived a
$1/d$ expansion of $\beta_{MM}^{RL}(d)$. One way to do this is to set
$\tilde{G}_d(x)=G_d({\tilde{x}=x/d+1/2})$ and then write $\tilde{G}_d(x)$ as
a power series in $1/d$. From this we find~\cite{Boutet96}
\begin{multline}
\beta_{MM}^{RL}(d)=\frac{D_1(d)}2\\
\times\left[1+\frac{1-\gamma}d+
\frac{\pi^2/12+\gamma^2/2-\gamma}{d^2}+O\left(\frac1{d^3}\right)\right]
\label{EquRlDev}
\end{multline}
where $\gamma=0.5772\ldots\,$ is Euler's constant.

\subsection{Numerical verifications and a new conjecture}
\label{SectRLNum}

Brunetti {\em et al.}~\cite{BrunettiKrauth91} have used numerical
simulations of the random-link models to confirm the predictions of
$\beta^{RL}$ to the level of 0.2\% for the MMP and 0.7\% for the MBMP at
$d=1$ and $d=2$. They have also checked the $O(1/N)$ corrections to
$\langle L^{RL}\rangle$ at $d=1$ and find relatively good agreement with the
theory.

To obtain further confirmation we have estimated $\beta_{MM}^{RL}(d)$ and
$\beta_{MBM}^{RL}(d)$ numerically for $1\leq d\leq 10$, and have found
accordance with the replica symmetric predictions to the level of 0.03\% for
the MMP and to the level of 0.1\% for the MBMP. In order to reduce the
statistical fluctuations and get quantitative errors on our estimates for
$\beta$, we use the following procedure. First we
compute the ensemble average $\langle L_{MM}^{RL}\rangle/N^{1-1/d}$ using a
variance reduction trick~\cite{CerfBoutet97,GrassbergerFreund90}. 
(The values of $N$ used here and for the other calculations
of $\beta$ given later on are 
$N=50, 70, 100, 150, 200, 260$, and $400$; we performed
averages over $5 \cdot 10^5$ instances at 
$N=50$ down to $7,500$ instances for
$N=400$.)
Then, to get the large $N$
limit, we fit our data using a $1/N$ series (truncated after the second order)
as indicated by the theory (equation~\ref{EquNDev}). The fits are good, with
$\chi^2$ values confirming the finite-size scaling law. The statistical
error bar on $\beta_{MM}^{RL}(d)$ is then obtained by the standard
method~\cite{BevingtonRobinson94}, whereby fixing the fit's leading coefficient
at $\beta\pm\sigma$ makes $\chi^2$ increase by one from its minimum value.
Our results are summarized in
Table~\ref{TableBetaRL}.

\begin{table}
\caption{Comparison of theoretical and numerical values of
$\beta_{MM}^{RL}(d)$ and $\beta_{MBM}^{RL}(d)$. Numbers in parentheses 
represent the statistical error bar
on the last digit(s).}
\label{TableBetaRL}
\begin{center}
\begin{tabular}{|c|ccc|}
\hline\noalign{\smallskip}
& & MMP & MBMP\\
$d$ & $\beta_{th}$ & $\beta_{num}$ & $\beta_{num}/2$\\
\noalign{\smallskip}\hline\noalign{\smallskip}
1 & 0.411234 & 0.41142(12) & 0.41134(9)\\
2 & 0.322580 & 0.32257(5) & 0.32262(4)\\
3 & 0.326839 & 0.32684(4) & 0.32691(4)\\
4 & 0.343227 & 0.34323(3) & 0.34327(4)\\
5 & 0.362175 & 0.36210(3) & 0.36222(3)\\
6 & 0.381417 & 0.38143(3) & 0.38143(3)\\
7 & 0.400277 & 0.40026(5) & 0.4002(5)\\
8 & 0.418548 & 0.41852(5) & 0.4180(5)\\
9 & 0.436185 & 0.43612(5) & 0.4362(5)\\
10 & 0.453200 & 0.45310(5) & 0.4531(4)\\
\noalign{\smallskip}\hline
\end{tabular}
\end{center}
\end{table}

We have also checked the prediction for the distribution $P_d(l)$ of the
bond lengths in the minimum (bipartite) matching. We find such good
agreement with theory that the numerical data is indistinguishable from the
replica predictions (on a figure one would not be able to tell
the two curves apart).

We will now use $P_d(l)$ to obtain an
estimate of the quantities computed numerically in Table~\ref{TableGaussRL}.
Neglecting the correlations among the bond lengths in the optimal matching,
we can use equation~\ref{EquPl} to predict $\sigma$ and $s$ as a function of
$N$. At $d=1$, we find $\sigma\sqrt{N/2}\approx 0.73$ and
$s\sqrt{N/2}\approx 1.85$ for the MMP. For the MBMP, the same formulae apply
if $N/2$ is replaced by $N$. These theoretical predictions (to be compared
with Table~\ref{TableGaussRL}) are about 10\% too small for $\sigma$ and
about 30\% too large for $s$, showing that correlations cannot be neglected,
but are nevertheless relatively small.

Another interesting quantity is the mean fraction of points connected to
their $k$th nearest neighbor in the optimal matching. Call this fraction
$p_d(k)$. In view of our numerical data at $d=1$ (see Table~\ref{TablePk}),
we conjecture for both the MMP and the MBMP that 
in the limit $N\rightarrow\infty$,
\begin{equation}
p_1(k)=2^{-k}.
\end{equation}
In addition the $N$-dependence of these fractions seems to be linear in
$1/N$ as one could expect from equation~\ref{EquNDev}. It seems likely that
our conjecture may be confirmed by using the replica method. There
may also be analogous relations in higher dimensions, but unfortunately
we have not found any convincing formulae. At best, 
our data are approximately fitted by stretched exponentials
(see also \cite{GrassbergerFreund90}).
This same kind of behavior also arises in other
combinatorial optimization problems
such as the traveling salesman problem~\cite{PercusMartin98}.
It is also interesting to note that another
conjecture has been proposed for a random-link MBMP by
Parisi~\cite{Parisi98}.  His conjectured relation gives
the mean length of the matching at all values of $N$, not just in the
limit $N \to \infty$.

\begin{table*}
\caption{Numerical results of $10^5(p_d(k)-2^{-k})$ for the
random-link MMP and MBMP in the case $d=1$. Numbers in parentheses
represent the statistical error on the last digit(s).}
\label{TablePk}
\begin{center}
\begin{tabular}{|c|ccc|ccc|}
\hline\noalign{\smallskip}
& \multicolumn{3}{c|}{MMP} & \multicolumn{3}{c|}{MBMP}\\
$k$ & $N=50$ & $N=100$ & $N=200$ & $N=50$ & $N=100$ & $N=200$\\
\noalign{\smallskip}\hline\noalign{\smallskip}
1 & 422(22) & 181(16) & 86(16) & 1185(16) & 608(11) & 305(11)\\
2 & 174(19) & 96(14) & 47(14) & 113(14) & 47(10)& 17(10)\\
3 & -26(15) & -12(10) & 1(10) & -238(10) & -108(7) & -52(7) \\
4 & -81(11) & -30(8) & -18(8) & -289(7) & -157(5) & -77(5)\\
5 & -112(8) & -49(5)& -31(5) & -248(5) & -123(4) & -56(4)\\
\noalign{\smallskip}\hline
\end{tabular}
\end{center}
\end{table*}

\section{Euclidean models}
\label{EUModels}

\subsection{The large $N$ limit}
\label{SectEUMdl}
Let $L_{MM}^E$ be the length of the minimum matching in the Euclidean MMP.
Hereafter, we take the points to be distributed randomly in the
$d$-dimensional unit hypercube.
Following the same argument as for the random-link MMP, one expects
$L_{MM}^{E}$ to scale as $N^{1-1/d}$. In fact, it has been
proven~\cite{Steele97} that the Euclidean MMP has the self-averaging
property in any dimension, so $L_{MM}^{E}/N^{1-1/d}$ tends to a constant
$\beta_{MM}^E(d)$ as $N\rightarrow\infty$ with probability one.

For the Euclidean MBMP, the situation is more complex. There are two
sets of points, so local density differences have a large 
effect. In particular, the
$N^{1-1/d}$ scaling law is {\em not} valid for $d\leq 2$. At $d=1$, it is
easy to see that the optimum corresponds to matching the points left to
right. Then a quick estimate shows that $L_{MBM}^{E}$ scales as $\sqrt N$
instead of as $N^0$. Furthermore $L_{MBM}^{E}/\sqrt N$ does not become
peaked, so there is no self-averaging. At $d=2$, the situation is more
interesting: $L_{MBM}^{E}$ scales~\cite{AjtaiKomlos84} as $\sqrt{N\ln N}$
(instead of as $\sqrt N$ for $L_{MBM}^{RL}$). The question of self-averaging
has not yet been settled but numerical simulations indicate that
the property does hold, and we will assume this is the case hereafter. 
At higher dimensions, we have proven self-averaging (see
Appendix~\ref{StrongSA}), so the quantity
$\displaystyle\beta_{MBM}^E(d)=\lim_{N\rightarrow\infty}L_{MBM}^E/N^{1-1/d}$
exists for $d\geq 3$.

As with the random-link models, one may wonder whether the central limit
theorem is at work. Following what was presented in
section~\ref{SectRLLargeN}, it is natural to investigate the limiting
distribution of the optimum length. It is convenient in the numerical
study to avoid
boundary effects; to do so, we work in the unit hypercube with periodic
boundary conditions. Using numerical measurements of the moments of the
distributions we find that the MMP obeys the CLT scaling laws (see
Table~\ref{TableGaussEu}). This is not surprising: for the Euclidean MMP, we
can divide the hypercube into subvolumes. Then the length of the minimum
matching is close to the sum of the minimum matching length in these
subvolumes, and the CLT scalings should hold. On the other hand, the CLT
scalings do {\it not} hold for the MBMP, and furthermore the limit
distribution is {\em not} Gaussian. In particular, we find that $s$ does 
not tend to zero, rather it {\it grows} with $N$. 
The CLT argument used for the MMP does not apply to this problem because
the subvolumes just mentioned will not in general contain 
equal number of points from each set.
Nevertheless, as the dimension increases, the density fluctuations decrease,
explaining why at fixed $N$ the MBMP values of $\sigma$ and $s$ get closer
to the MMP ones as $d$ increases. (For a formal application of this
argument, see Appendix~\ref{StrongSA}.)

\begin{table*}
\caption{Numerical results for the relative standard deviation $\sigma$
and skewness $s$ of the distribution of $L_{MM}^E$ and
$L_{MBM}^E$ at $d=2$, 3 and 4.}
\label{TableGaussEu}
\begin{center}
\begin{tabular}{|c|cc|cc|cc|}
\hline\noalign{\smallskip}
& \multicolumn{2}{c|}{MMP $d=2$} & \multicolumn{2}{c|}{MMP $d=3$} &
\multicolumn{2}{c|}{MMP $d=4$}\\
$N$ & $\sigma\sqrt{N/2}$ & $s\sqrt{N/2}$ & $\sigma\sqrt{N/2}$ &
$s\sqrt{N/2}$ & $\sigma\sqrt{N/2}$ & $s\sqrt{N/2}$\\
\noalign{\smallskip}\hline\noalign{\smallskip}
50 & 0.302 & -0.90 & 0.244 & -0.65 & 0.206 & -0.68\\
100 & 0.299 & -0.93 & 0.244 & -0.66 & 0.205 & -0.64\\
200 & 0.290 & -0.87 & 0.243 & -0.52 & 0.204 & -0.52\\
400 & 0.295 & -0.85 & 0.243 & -0.88 & 0.203 & -0.45\\
\noalign{\smallskip}\hline\hline\noalign{\smallskip}
& \multicolumn{2}{c|}{MBMP $d=2$} & \multicolumn{2}{c|}{MBMP $d=3$} &
\multicolumn{2}{c|}{MBMP $d=4$}\\
$N$ & $\sigma\sqrt{N}$ & $s\sqrt{N}$ & $\sigma\sqrt{N}$ & $s\sqrt{N}$ &
$\sigma\sqrt{N}$ & $s\sqrt{N}$\\
\noalign{\smallskip}\hline\noalign{\smallskip}
25 & 0.603 & 2.61 & 0.350 & 1.16 & 0.258 & 0.42\\
50 & 0.744 & 4.20 & 0.387 & 1.94 & 0.274 & 1.12\\
100 & 0.938 & 6.49 & 0.431 & 3.26 & 0.289 & 1.88\\
200 & 1.197 & 9.97 & 0.481 & 4.71 & 0.305 & 2.53\\
\noalign{\smallskip}\hline
\end{tabular}
\end{center}
\end{table*}

\subsection{Numerical results and the random-link approximation}
\label{SectEUNum}
To date, little has been done to compute the ground-state energy densities
$\beta_{MM}^E(d)$ and $\beta_{MBM}^E(d)$. The best estimates prior to our
recent work~\cite{BoutetMartin97} were
\cite{Smith89,GrassbergerFreund90} $\beta_{MM}^E(2)\approx0.312$ and
$\beta_{MM}^E(3)\approx0.318$; for $\beta_{MBM}^E(d)$, no valid
estimates have yet been published.  Expanding upon the work
in~\cite{BoutetMartin97}, we
now provide very accurate measurements of these quantities,
using the same procedure
as in the random-link case. 
Again we find that the
$\chi^2$ values justify the use of a truncated $1/N$ series.

\begin{table*}
\caption{Comparison of MMP ground state energies for the three models:
Euclidean, random-link, and random-link including 3-link Euclidean
corrections ($1\leq d\leq 10$). For $\beta^E(d)$ the numbers in parentheses
are statistical errors on the last digit.}
\label{TableBetaMM}
\begin{center}
\begin{tabular}{|c|cccccc|}
\hline\noalign{\smallskip}
$d$ & $\beta^E(d)$ & $\beta^{RL}(d)$ & $\frac{\beta^{RL}-\beta^E}{\beta^E}$ &
$d\frac{\beta^{RL}-\beta^E}{\beta^E}$ &
$\beta^{EC}(d)$ & $\frac{\beta^{EC}-\beta^E}{\beta^E}$\\
\noalign{\smallskip}\hline\noalign{\smallskip}
1 & 0.5 & 0.411234 & -17.75\% & -0.178 &&\\
2 & 0.3104(2) & 0.322580 & 3.92\% & 0.078 & 0.30915 & -0.40\%\\
3 & 0.3172(2) & 0.326839 & 3.04\% & 0.091 & 0.31826 & 0.33\%\\
4 & 0.3365(3) & 0.343227 & 2.01\% & 0.080 & 0.33756 & 0.30\%\\
5 & 0.3572(2) & 0.362175 & 1.39\% & 0.070 & 0.35818 & 0.27\%\\
6 & 0.3777(1) & 0.381417 & 0.98\% & 0.059 & 0.37849 & 0.21\%\\
7 & 0.3972(1) & 0.400277 & 0.77\% & 0.054 & 0.39807 & 0.22\%\\
8 & 0.4162(1) & 0.418548 & 0.56\% & 0.045 & 0.41685 & 0.17\%\\
9 & 0.4341(1) & 0.436185 & 0.48\% & 0.042 & 0.43485 & 0.17\%\\
10 & 0.4515(1) & 0.453200 & 0.38\% & 0.038 & 0.45214 & 0.14\%\\
\noalign{\smallskip}\hline
\end{tabular}
\end{center}
\end{table*}

\begin{table}
\caption{Comparison of MBMP ground state energies for the two models:
Euclidean and random-link ($3\leq d\leq 10$). The rightmost column compares
the Euclidean MBMP and MMP models. For $\beta^E(d)$ the numbers in
parentheses are statistical errors on the last digit(s). In the last column
$\Delta_E=(\beta_{MBM}^E-2\beta_{MM}^E)/\beta_{MBM}^E$.}
\label{TableBetaMBM}
\begin{center}
\begin{tabular}{|c|cccc|}
\hline\noalign{\smallskip}
$d$ & $\beta^E(d)$ & $\beta^{RL}(d)$ &
$\frac{\beta^{RL}-\beta^E}{\beta^E}$ & $d^2\Delta_E$\\
\noalign{\smallskip}\hline\noalign{\smallskip}
3 & 0.7080(2) & 0.653679 & -7.68\% & 0.664\\
4 & 0.7081(2) & 0.686455 & -3.06\% & 0.597\\
5 & 0.7349(1) & 0.724350 & -1.44\% & 0.514\\
6 & 0.7688(3) & 0.762834 & -0.77\% & 0.482\\
7 & 0.8039(2) & 0.800554 & -0.41\% & 0.461\\
8 & 0.8391(2) & 0.837097 & -0.24\% & 0.445\\
9 & 0.8736(2) & 0.872370 & -0.14\% & 0.429\\
10 & 0.9076(2) & 0.906400 & -0.13\% & 0.436\\
\noalign{\smallskip}\hline
\end{tabular}
\end{center}
\end{table}

As previously remarked, our random-link distributions were established in
order to match the one- and two-link distributions of the Euclidean model.
So, if the effects coming from the Euclidean correlations among three or
more link lengths are small, then the properties of the random-link and
Euclidean M(B)MP should be quantitatively close. In fact, replacing
$\beta_{MM}^E(d)$ and $\beta_{MBM}^E(d)$ by $\beta_{MM}^{RL}(d)$ and
$\beta_{MBM}^{RL}(d)$ leads to a very precise approximation. As shown in
Table~\ref{TableBetaMM}, $\beta_{MM}^{RL}(d)$ differs from $\beta_{MM}^E(d)$
by 17.8\% at $d=1$ and by 3.9\% at $d=2$, and this difference decreases
quickly as the dimension increases (the quantity $\beta^{EC}$ given in the
table will be discussed in section~\ref{SectCorrect}). Likewise for the MBMP,
shown in Table~\ref{TableBetaMBM}: $\beta_{MBM}^{RL}(d)$ differs from
$\beta_{MBM}^E(d)$ by only 7.7\% at $d=3$. Note that comparing
$\beta_{MBM}^{RL}(d)$ and $\beta_{MBM}^E(d)$ at $d\leq 2$ is meaningless
since the scaling laws are different as we mentioned in the beginning of
section \ref{SectEUMdl}. Nevertheless, we have also 
computed $\beta_{MBM}^E(2)$;
empirically, we found that using a $1/N$ fit could not do,
but that good $\chi^2$ values were obtained using
a linear fit in $1/\ln(N)$. We find
$\beta_{MBM}^E(2)=0.340(1)$.

Since the random-link approximation was also found to be very good for the
traveling salesman problem \cite{CerfBoutet97}, our results
suggest that this approximation should be widely applicable to link-based
optimization problems. Furthermore, we can understand how the size of the
error inherent to this approximation decreases as $d\to\infty$. Consider for
instance the bond occupation probabilities for links connected to a given
site. One expects that these probabilities have a large $d$ limit, and that
Euclidean correlations introduce $1/d$ corrections compared to the
random-link case. (To be precise, one expects the corrections
to be given by a $1/d$ expansion, with
the leading term naturally being of order $1/d$.) In support of this
argument, we have checked that the relative difference between the Euclidean
value and the random-link value of $p_d(k)$ is indeed 
of order $1/d$. We can then see
how quantities such as $\beta(d)$ depend on the link lengths 
vary when one goes from the random-link model to 
the Euclidean model. At large
$d$, a simple calculation shows~\cite{CerfBoutet97} that the 
mean link lengths between $k$th nearest neighbors have for
different values of $k$ a relative difference 
of $O(1/d)$ (regardless of whether we use the random-link
or the Euclidean ensemble).  Since the
{\it occupation probabilities} $p_d(k)$
have Euclidean corrections of order $1/d$, the random-link
approximation leads to an error of order $1/d^2$ for $\beta(d)$ and
$P_d(l)$. We have tested this behavior numerically by considering the quantity
$d(\beta_{MM}^{RL}-\beta_{MM}^E)/\beta_{MM}^E$, and the data
scale as $1/d$ as expected. Thus the random-link approximation gives both the leading
and subleading $1/d$ dependence of $\beta_{MM}^E(d)$. Then, from
equation~\ref{EquRlDev}, we find
\begin{equation}
\beta_{MM}^E(d)=\frac{D_1(d)}2\left[1+
\frac{1-\gamma}d+O\left(\frac1{d^2}\right)\right].
\label{EquEuDev}
\end{equation}
Furthermore we have directly confirmed this dependence by performing a fit of
our $\beta_{MM}^E(d)$ values and we find $0.424\pm0.008$ for the coefficient
of the $1/d$ term; this is to be compared to the theoretical value
$1-\gamma=0.42278\ldots$~.

Performing the same analysis for the MBMP, we find that $\beta_{MBM}^E(d)$
also satisfies Equation~\ref{EquEuDev} (omitting the factor $1/2$) or
equivalently that $(\beta_{MBM}^E-2\beta_{MM}^E)/\beta_{MBM}^E=O(1/d^2)$ (see
Table~\ref{TableBetaMBM}). A direct fit to the $1/d$ term of
Equation~\ref{EquEuDev} for $\beta_{MBM}^E$ gives $0.42\pm0.05$ in agreement
with the theoretical value $1-\gamma$.

\begin{figure}
\begin{center}
\resizebox{0.45\textwidth}{!}{\includegraphics{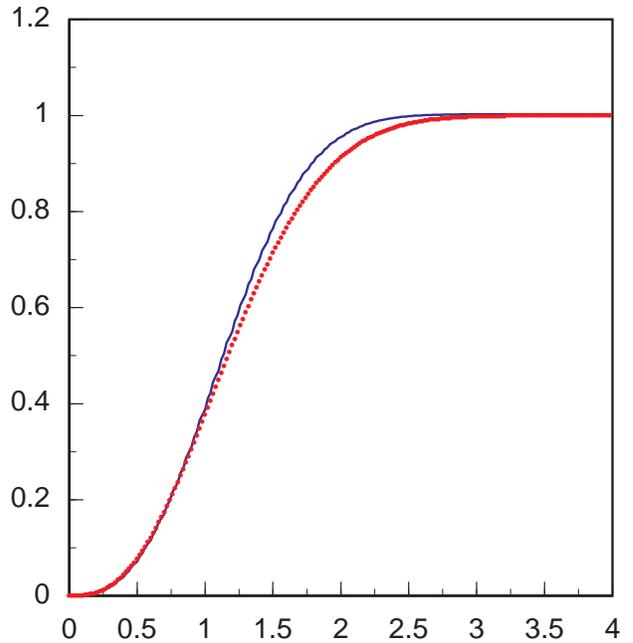}}
\end{center}
\caption{Comparison of the random-link (continuous line) and Euclidean
(points) integrated distributions of the rescaled bond lengths in the minimum
bipartite matching for the MBMP at $d=3$ and $N=60$.}
\label{FigLink}
\end{figure}

Finally, we have also applied the random-link approximation to compare $P_d(l)$
in the random-link and Euclidean models and found very good agreement (see
Figure~\ref{FigLink} for the MBMP at $d=3$, which has the largest
discrepancy). The agreement becomes better as the dimension increases, with
an error of order $1/d^2$.

\subsection{Euclidean corrections to the random-link approximation}
\label{SectCorrect}
The mean field model ({\it i.e.}, the random-link model) provides
a very good approximation to the finite dimensional Euclidean model.
It is 
of major interest to push the approximation further and to derive, for
instance, a large dimensional expansion. 
For the MMP, M \& P have calculated~\cite{MezardParisi88} a
correction to the random-link approximation by considering the
effects of the three-link Euclidean correlations. Such correlations arise
only when three links form a triangle. (For the MBMP, one would have to go
to four-link correlation effects, and this has not yet been attempted). M \&
P's result is given in terms of a function $G_d$, but where $G_d$ now
satisfies a much more complicated integral equation (equation 34 in their
paper). From this, one obtains a new estimate for $\beta_{MM}^E(d)$,
which we denote here by $\beta_{MM}^{EC}(d)$ (EC stands for Euclidean
corrections).

We solved numerically this modified integral equation for $G_d$ and
computed $\beta_{MM}^{EC}(d)$ for $2\leq d\leq 10$ (see
Table~\ref{TableBetaMM}). Comparing with $\beta_{MM}^{E}(d)$,
we see that these estimates are considerably more
accurate than when using $\beta_{MM}^{RL}(d)$. At $d=2$, the 
random-link approximation leads to an error of
3.9\%; this error is reduced by nearly a factor of 10 by including the
corrections due to 3-link correlations. At $d=3$, the error is
reduced from 3.0\% to 0.4\%. At larger $d$, the error continues to decrease,
though the effect is less significant.

To understand the dependence of these corrections on dimension, it 
is useful to consider how the difference
$\beta_{MM}^{EC}-\beta_{MM}^{RL}$ scales with $d$. This difference is
associated with a $3$-link correction term which gives the probability of
finding nearly equilateral triangles as $d\rightarrow\infty$. It is not
difficult to see that this probability goes to zero exponentially with $d$.
Thus the $3$-link correlations give tiny corrections at large $d$ (as
confirmed by the numerics), and the power series expansion in $1/d$ of
$\beta_{MM}^{EC}$ is {\em identical} to that of $\beta_{MM}^{RL}$. In fact
the $1/d$ series of $\beta_{MM}^{EC}$ is not modified if 
one includes 4, 5, or {\it any}
finite number of multi-link correlations. This follows from the fact that
all non-zero multi-link correlations come from
sets of links forming at least one loop, and that
fixed sized ($N$-independent) loops connecting near neighbors become
exponentially rare at large $d$. (Of course this behavior depends on having
randomly placed points; the situation is very different 
on lattices, for instance, where small loops are important.)
The main consequence of the nature of these multi-link correlations
is that the 
M \& P $k$-link correlation
expansion will not converge towards $\beta_{MM}^E(d)$, and in 
particular it does not even allow one to compute the $O(1/d^2)$ term
in the $1/d$ expansion of $\beta_{MM}^E(d)$. Formally, the
Euclidean dimensional dependence is beyond all orders 
of the $k$-link
correlation expansion. This non-perturbative behavior is quite 
remarkable, and indicates that the
large $N$ limit and the $k$-link correlation expansion do not commute. 
This property may have its analogue in other disordered
systems.

\section{Summary and Discussion}
\label{Discussion}
In this article we have studied two versions of the stochastic minimum
matching and minimum bipartite matching problems: the random-point Euclidean
ensemble, and its correlation-free approximation, the random-link ensemble.
For both ensembles, we have given evidence that the ground state energy is
self-averaging and obeys (except for the Euclidean MBMP) the central limit
theorem scaling laws. For both ensembles we have performed extensive
numerical simulations in order to measure the ground state energy density
$\beta(d)$ and the distribution $P_d(l)$ of bond lengths in the
ground-state. For the random-link model we have checked to high precision
the replica symmetric prediction of M\'ezard and Parisi and find 
excellent agreement. Furthermore, we have proposed
a new conjecture at $d=1$ that
suggests further analytical calculations. For the Euclidean model, we have
studied the accuracy of the random-link approximation and find the error to
be small even at low dimensions. For example at $d=2$ the error introduced
by this approximation is 3.9\% for $\beta_{MM}^E$. We have also been able to
go beyond the random-link approximation by applying 
the formalism of M\'ezard and
Parisi in order to include 3-link correlations associated with the triangle
inequality. We find that this improved estimate reduces the error by nearly
a factor of 10 at low dimensions. In particular, the resulting $d=2$
prediction for $\beta_{MM}^E$ has an error of 0.4\%.

The limit of high dimensions is also of major interest. Based on
our simulations up to $d=10$, 
we have given strong evidence that the first two terms of the $1/d$
expansion of the random-link model are the same as for the Euclidean model
but that these two models differ at order $1/d^2$. Furthermore we have
argued that for any fixed $k$, $k$-link correlations in Euclidean space only
give rise to exponentially small contributions in $d$ and thus do not modify
the $1/d$ expansion. As a consequence, the Euclidean 
dimensional dependence is beyond
all orders of the $k$-link correlation expansion. 

Although our study was performed in the context of the MMP, our reasoning
applies also to other link-based problems associated with random points, and
leads us to the following picture. For these random-point systems, whenever
the thermodynamic functions depend only on the {\em local} properties of the
(short) link graph, we expect the error in the random-link approximation to
be exponentially small in $d$. This will always be the case in the high
temperature phase where correlations are weak. However, in the
low temperature phase, the correlations may be such that the
$N\rightarrow\infty$ limit and the $k$-link correlation expansion do not
commute. We expect this to be the case in many combinatorial optimization
problems such as the assignment problem and the traveling salesman problem,
where $k$-link correlations with $k$ growing in $N$ remain important as
$N\rightarrow\infty$.  Arbitrarily large loops matter in these systems, and
contribute to the thermodynamics at order $1/d^2$. This change in behavior
can be illustrated using more physical language by considering a polymer on
a random-point lattice. Then the arguments given previously indicate that the
random-link approximation is exponentially good in the ``dilute'' phase,
while it leads to an error in powers of $1/d$ in the collapsed phase. This
non-perturbative behavior of the $1/d$ expansion is reminiscent of what
occurs in lattice systems such as the Eden model, where the $1/d$ expansion
does not commute with the $N\rightarrow\infty$ limit~\cite{Friedberg85}. We
would not be surprised if similar phenomena occurred in other disordered
systems, whether they be based on links or spins; the calculation of their
$1/d$ series would then be particularly difficult. Further insight into
these issues may be obtained by looking at excited states of the M(B)MP to
see whether they consist of arbitrary large loops (i.e., whose sizes diverge
as $N\rightarrow\infty$). Such a study could also allow a direct
investigation of possible replica symmetry breaking in the Euclidean case.

It is tempting to speculate on how all of this might carry over to
spin-glasses. There, one analogue of the random-point MMP is the
Edwards-Anderson model with nearest neighbor interactions on a
$d$-dimensional hypercubic lattice of connectivity $2 d$. The mean field
model for spin-glasses is usually taken to be the Sherrington-Kirkpatrick
infinite connectivity model. However, a more appropriate mean field model in
this context is that of Viana and Bray \cite{VianaBray85} in which the
connectivity is $2 d$ and for which the couplings $J_{ij}$ have the same
{\it individual} distribution as in the Edwards-Anderson model. Just as in
the matching problem, one thereby obtains a mean field model at any given
dimension, and the mean field approximation then consists of using this
model to estimate the thermodynamic quantities of the $d$-dimensional
Edwards-Anderson model. We expect this approach to lead to errors of a few
percent at low dimensions, and to $O(1/d)$ errors at large $d$. Unlike the
random-link MMP, the Viana-Bray model has not yet been solved analytically,
so that the mean field values would have to be calculated approximately.
Nevertheless, 
we view these mean field models as providing a very promising approach
to computing quantities in the Edwards-Anderson model. We hope that
the potential reward will encourage new attempts to solve the
Viana and Bray model, and that the challenge of determining the 
first $O(1/d)$ correction to ``mean field'' will be taken up.

\begin{acknowledgement}
We are grateful to D. Dean,
W. Krauth, M. M\'ezard, H. Orland, 
N. Sourlas and J. Vannimenus for comments, and to O. Bohigas
for his constant interest. Special thanks go to A. Percus
for constructive criticisms.
J.H. and J.H.B.dM. acknowledge a fellowship from
the MENESR, and O.C.M. acknowledges support from the Institut Universitaire
de France. The Division de Physique Th\'eorique is an Unit\'e de Recherche
des Universit\'es Paris~XI et Paris~VI associ\'ee au CNRS.
\end{acknowledgement}

\appendix

\section{Bounds for the MMP}
\label{Bound}
Here we review exact lower and upper bounds for $\beta_{MM}^{RL}(d)$, and
provide to our knowledge the first finite upper bound at $d=1$. First there
is a trivial lower bound: each point is at best linked to its nearest
neighbor, so
\begin{equation}
\beta_{MM}^{RL}(d)\geq D_1(d)/2.
\end{equation}
To get an upper bound, we use another optimization problem: the traveling
salesman problem (TSP), which consists of finding a minimum length tour
visiting all the $N$ points. Indeed we can obtain a matching by removing
every second bond in a tour (note that this argument fails for the MBMP), so
we get
\begin{equation}
\beta_{MM}^{RL}(d)\leq\beta_{TSP}^{RL}(d)/2,
\end{equation}
where $\beta_{TSP}^{RL}(d)$ is the analog of $\beta_{MM}^{RL}(d)$ for the
TSP. A bound for $\beta_{TSP}^{RL}(d)$ is already known for $d\geq 2$ (see
in~\cite{VannimenusMezard84}). This bound comes from the greedy algorithm
and is
\begin{equation}
\beta_{TSP}^{RL}(d)\leq\frac{D_1(d)}{1-1/d}.
\end{equation}
In the case $d=1$, the greedy construction leads to a bound which grows
logarithmically in $N$ and thus is useless. We have obtained a finite bound
for the case $d=1$ using a different approach. The idea is that a bound is
known~\cite{Karp79,KarpSteele85} for the {\em asymmetric} TSP (where
$l_{ij}$ and $l_{ji}$ are independent). Let us call $\lambda_{ij}$ the link
lengths of the asymmetric TSP. We denote by $\rho_{Asym}(l)$ the distribution
density of the $\lambda_{ij}$, and associate a symmetric TSP to any
asymmetric TSP by setting $l_{ij}=\min(\lambda_{ij},\lambda_{ji})$. Then, in
the limit of short link lengths, the distribution in this symmetric TSP is
$\rho_{Sym}(l)=2\rho_{Asym}(l)$. This gives
\begin{equation}
\beta_{TSP}^{Sym.}(d)\leq2^{1/d}\beta_{TSP}^{Asym.}(d).
\end{equation}
And we thus obtain our bound at $d=1$.

\section{Self-averaging for the Euclidean MBMP}
\label{StrongSA}
We shall denote by $L_{MBM}(X_1,\ldots,X_N,Y_1,\ldots,Y_N)$ the length of
the minimum bipartite matching between the $X_i$'s and the $Y_i$'s. We prove
here that for any $d\geq3$, $L_{MBM}$ satisfies a theorem analogous to the
one Beardwood, Halton and Hammersley have proven for the
TSP~\cite{BeardwoodHalton59}. Specifically, we prove the following:

{\em Let $X_1,\ldots,X_N,\ldots$ and $Y_1,\ldots,Y_N,\ldots$ be two
sequences of random points independently and uniformly distributed in
$[0,1]^d$, where $d\geq 3$, and let
$L_N=L_{MBM}(X_1,\ldots,X_N;Y_1,\ldots,Y_N)$. There exists a constant
$\beta_{MBM}(d)>0$ such that with probability one,}
\begin{equation}
\lim_{N\rightarrow\infty}\frac{L_N}{N^{1-1/d}}=\beta_{MBM}(d).
\end{equation}

To begin with,
we remark that to prove this theorem, it is sufficient to establish
that $L_N/N^{1-1/d}$ converges in mean value to a constant $\beta_{MBM}(d)$.
This is a consequence of the following lemma~\cite{Talagrand92}: {\em For
any $t>0$,}
\begin{equation}
P\left(\left|\frac{L_N}{N^{1-1/d}}-
\langle\frac{L_N}{N^{1-1/d}}\rangle\right|>t\right)\leq2\exp(-\frac{N^{1-2/d}t^2}{8d}).
\end{equation}
The theorem then follows easily from the convergence of $\langle
L_N\rangle/N^{1-1/d}$ as $N\rightarrow\infty$, by applying the
Borel-Cantelli lemma. We now wish to establish that for $d\geq 3$ the
quantity $\langle L_N\rangle/N^{1-1/d}$ indeed converges to a constant
$\beta_{MBM}(d)>0$. To do this, we exploit the subadditivity properties of
$L_{MBM}$ (see~\cite{Steele81}).

First we need to generalize $L_{MBM}$ to matchings between two sets of
different cardinalities. We shall define
$L_{MBM}(X_1,\ldots,X_{N_1};Y_1,\ldots,Y_{N_2})$ by requiring that the
matchings contain as few unmatched points as possible, that is we leave
$|N_1-N_2|$ points unmatched.

Suppose the points $X_1,\ldots X_{N_1},Y_1,\ldots Y_{N_2}$ belong to an
arbitrary cube $Q$ whose edges have length $a$, and divide $Q$ into disjoint
cubes $Q_p,~p=1,\ldots,2^d$ by splitting each edge in two halves. Construct
in each $Q_p$ an optimal matching in the sense just defined, between the
$n_{1,p}$ points $X_i$ and the $n_{2,p}$ points $Y_i$ in $Q_p$, and denote
its length by $L_p$. There will be $|n_{1,p}-n_{2,p}|$ points left
unpaired in each $Q_p$, so if $L_0$ denotes the length of an
optimal matching for these points, one has
\begin{multline}
\label{EquDecim}
L_{MBM}(X_1,\ldots,X_{N_1};Y_1,\ldots,Y_{N_2})\leq 
\sum_{p=1}^{2^d} L_p + L_0\\
\leq \sum_{p=1}^{2^d} L_p + \frac12 a\sqrt d 
\sum_{p=1}^{2^d} |n_{1,p}-n_{2,p}|.
\end{multline}

Now we apply this to $Q=[0,1]^d$. Let $Q_{p_1}, p_1=1,\ldots 2^d$ be the
cubes obtained in the last subdivision, let $Q_{p_1p_2}$ be the cubes
obtained by splitting in two halves the edges of each $Q_{p_1}$, and so on.
By repeating this operation $K$ times, we get a subdivision with cubes
$Q_{p_1\ldots p_K}$ whose edges are of length $1/2^K$. Let $n_{1,p_1\ldots
p_K}$ and $n_{2,p_1\ldots p_K}$ be respectively the number of points $X_i$
and $Y_i$ in $Q_{p_1\ldots p_K}$. Apply (\ref{EquDecim}) first to the
$Q_{p_1,\ldots p_{K-1}}$'s, then to the $Q_{p_1\ldots p_{K-2}}$'s, etc.,
keeping at each step only those points that are still unpaired. It is easy
to convince oneself that the number of unpaired points in each
$Q_{p_1,\ldots,p_{K-k}}$ just after step $k$ is given by
$|n_{1,p_1,\ldots,p_{K-k}}-n_{2,p_1,\ldots,p_{K-k}}|$. After step $k=K$ one
obtains a matching between $X_1,\ldots,X_{N_1}$ and $Y_1,\ldots,Y_{N_2}$
where all but $|N_1-N_2|$ of the points are matched. One is thus led to the
following inequality:
\begin{multline} \label{EquSousAdd}
L_{MBM}(X_1,\ldots X_{N_1};Y_1,\ldots Y_{N_2}) 
\leq \sum_{p_1\ldots p_K} L_{p_1\ldots p_K}\\
+\sum_{k=1}^K \frac{\sqrt d}{2^k}
\sum_{p_1\ldots p_k} |n_{1,p_1\ldots p_k}-n_{2,p_1\ldots p_k}|.
\end{multline}
We now proceed to derive a subadditivity property for the average value of
$L_{MBM}$. To do this, it is useful to consider the case where $N_1$ and
$N_2$ are not fixed integers but are independent Poisson random variables
with the same fixed parameter $N$. For a given $k$, the numbers
$n_{1,p_1,\ldots p_k}$ and $n_{2,p_1,\ldots p_k}$ are then also independent
Poisson random variables, with parameter $N/2^{kd}$. Let
$M(N)=\langle L_{MBM}(X_1,\ldots,X_{N_1};Y_1,\ldots,Y_{N_2})\rangle$. It is
easy to see that
\begin{equation}
\langle L_{p_1\ldots p_K}\rangle = 2^{-K} M(N/2^{Kd}).
\end{equation}
Moreover, well known properties of Poisson variables allow us to write
\begin{equation}
\langle|n_{1,p_1\ldots p_k}-n_{2,p_1\ldots p_k}|\rangle \leq 
\sqrt 2 \left( \frac N{2^{kd}} \right)^{1/2}.
\end{equation}
By taking mean values in (\ref{EquSousAdd}), we are thus led to
\begin{equation}
M(N) \leq 2^{K(d-1)}M(N/2^{Kd}) + \sqrt{2dN} \sum_{k=1}^K 2^{k(d/2-1)}.
\end{equation}
Extending this construction, one may easily prove~\cite{Boutet96} that for
$2^K\leq m<2^{K+1}$ we have
\begin{equation}
M(N) \leq m^{d-1}M(N/m^d) + 2^d\sqrt{2dN} \sum_{k=0}^K 2^{k(d/2-1)}.
\end{equation}
Dividing this last inequality by $N^{1-1/d}$ and then replacing $N$ by
$m^dN$,
\begin{equation}
\frac{M(m^dN)}{(m^dN)^{1-1/d}} \leq \frac{M(N)}{N^{1-1/d}} + 
\frac{2^d\sqrt{2d}}{N^{1/2-1/d}} \sum_{k=0}^K 2^{-k(d/2-1)}.
\end{equation}
Standard arguments may now be used to show that the ratio $M(N)/N^{1-1/d}$
necessarily converges to a limit $\beta_{MBM}(d)$ as $N\rightarrow \infty$.
Indeed, let $f(t) = M(t^d)/t^{d-1}$. One verifies at once that $f(t)$
satisfies
\begin{equation} \label{EquFIneq}
f(mt)\leq f(t)+C_d/t^{d/2-1}
\end{equation}
for all $t>0$ and all integer $m$; $f(t)$ is continuous,
since $M(N)$ is a continuous function of $N$.
So the expression $f(t) + C_d/t^{d/2-1}$ is bounded in $[1,2]$ and since
$[1,\infty)$ is the union of the intervals $m[1,2], m\geq 1$, it follows
from (\ref{EquFIneq}) that $f(t)$ remains bounded as $t\rightarrow \infty$,
thus $\lim^* f(t) < \infty$. Now define $\beta=\lim_* f(t)$. For any
$\epsilon >0$, choose $t_0\gg 1$ and
$\eta >0$ such that $f(t)+C_d/t^{d/2-1} < \beta + \epsilon$
for $t$ in the interval $I=[t_0-\eta,t_0+\eta]$.
Since the intervals $mI$, $m\geq 1$ span a whole interval
$[A,\infty)$ for $A$ sufficiently large,
it follows again from (\ref{EquFIneq}) that
$\lim^* f(t)\leq \beta+\epsilon$.
Since $\epsilon$ is arbitrary, we have $\lim^* f(t)=\beta$, hence
$f(t) \rightarrow \beta$ as $t\rightarrow \infty$, from which it follows that 
$\lim_{N\rightarrow \infty} M(N)/N^{1-1/d}=\beta$.\qed

We have thus shown that for $d\geq 3$,
\begin{multline}
\langle L_{MBM}(X_1,\ldots,X_{N_1};Y_1,\ldots,Y_{N_2})\rangle\\
\sim \beta_{MBM}^E(d)N^{1-1/d},~N\rightarrow \infty
\end{multline}
when $N_1$ and $N_2$ are independent Poisson variables with parameter $N$.
The same result for the mean value $\langle L_N\rangle$, where $N$ is a fixed integer,
then follows easily. Indeed, one has the obvious bound
\begin{multline}
|L_{MBM}(X_1,X_N;Y_1,Y_N)-
L_{MBM}(X_1,X_{N_1};Y_1,Y_{N_2})|\\
\leq \sqrt d (|N_1-N|+|N_2-N|),
\end{multline}
and taking mean values and dividing by $N^{1-1/d}$, we deduce that 
\begin{equation}
\lim_{N\rightarrow \infty} \frac{\langle L_N\rangle}{N^{1-1/d}} = \beta_{MBM}(d).
\end{equation}
For further discussion of self-averaging proofs see \cite{Steele97,Azuma67}.

\bibliographystyle{unsrt}
\bibliography{/tmp_mnt/home/houdayer/Papers/Biblio/references}
\end{document}